# Theoretical Description of Hyperpolarization Formation in the SABRE-relay Method

**Stephan Knecht,[a] Danila A. Barskiy,[b] Gerd Buntkowsky,[a,*] Konstantin L. Ivanov[c,*]**

[a] *Eduard-Zintl Institute for Inorganic and Physical Chemistry, TU Darmstadt, Darmstadt 64287, Germany*

[b] *University of California at Berkeley, College of Chemistry and QB3, Berkeley, CA 94720, USA*

[c] *International Tomography Center, Siberian Branch of the Russian Academy of Sciences, and Novosibirsk State University, Novosibirsk, 630090, Russia*

* Corresponding authors, gerd.buntkowsky@chemie.tu-darmstadt.de and ivanov@tomo.nsc.ru

## Abstract

SABRE (Signal Amplification By Reversible Exchange) has become a widely used method for hyperpolarizing nuclear spins, thereby enhancing their Nuclear Magnetic Resonance (NMR) signals by orders of magnitude. In SABRE experiments, non-equilibrium spin order is transferred from parahydrogen to a substrate in a transient organometallic complex. Applicability of SABRE is expanded by the methodology of SABRE-relay, in which polarization can be relayed to a second substrate either by direct chemical exchange of hyperpolarized nuclei or by polarization transfer between two substrates in a second organometallic complex. To understand the mechanism of the polarization transfer and study the transfer efficiency, we propose a theoretical approach to SABRE-relay, which can treat both spin dynamics and chemical kinetics as well as the interplay between them. The approach is based on a set of equations for the spin density matrices of the spin systems involved (i.e., SABRE substrates and complexes), which can be solved numerically. Using this method, we perform a detailed study of polarization formation and analyse in detail the dependence of attainable polarization level on various chemical kinetic and spin dynamic parameters. We foresee applications of the present approach for optimizing SABRE-relay experiments with the ultimate goal of achieving maximal NMR signal enhancements for substrates of interest.

# I. Introduction

In less than a decade after its discovery,[1] the Signal Amplification By Reversible Exchange (SABRE) method[2, 3] has evolved into an established tool for enhancing weak Nuclear Magnetic Resonance (NMR) signals of various nuclei, such as $^{1}$H,[1-3] $^{13}$C,[1, 4-6] $^{15}$N,[4, 5, 7-9] $^{19}$F[10, 11] and $^{31}$P[12, 13]. Notably, SABRE can be used to boost the NMR signal of biologically important molecules such as antibiotics, vitamins[14, 15] and biorthogonal molecular tags[16]. In the SABRE method, strong non-equilibrium nuclear spin polarization of a suitable substrate, termed spin hyperpolarization, is generated. The source of this polarization is parahydrogen ($p$H$_2$), that is the hydrogen molecule in its nuclear spin singlet state. In SABRE experiments, the spin order of $p$H$_2$ is transferred to the substrate ($S_a$), in a transient complex ($C_1$), which simultaneously binds $S_a$ and $p$H$_2$. After dissociation of the complex, the generated hyperpolarization is accumulated in the free $S_a$ pool. The corresponding reaction scheme is shown in **Figure 1A**. A great advantage of the SABRE method compared to other hyperpolarization techniques is that the underlying chemical reactions are reversible and the polarization process can be repeated multiple times[17, 18] by flushing fresh $p$H$_2$ through the solution. As far as the mechanism of the polarization process is concerned, it usually relies on coherent spin dynamics; chemical components taking part in this process are indicated by arrows (**Figure 1A**). Suitable conditions for polarization transfer are generated at low fields (hyperpolarization of protons)[1, 19-21] or even at zero-to-ultralow-field conditions (namely, hyperpolarization of hetero-nuclei)[4, 5, 22-24]. Alternatively, one can use NMR radiofrequency (RF) pulses for generating hyperpolarization at high magnetic fields.[18, 25-32]

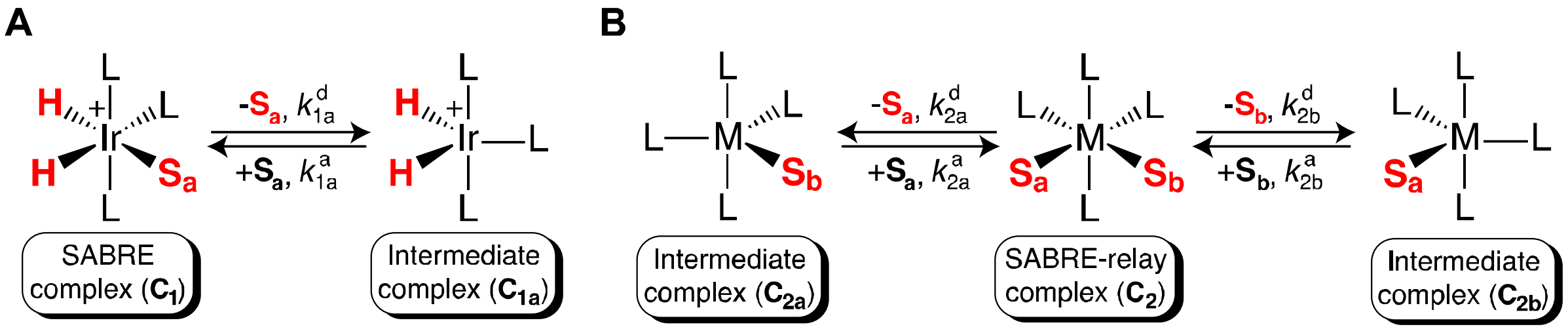

**Figure 1**: Schematic representation of the SABRE-relay process involving two organometallic complexes. A) Both the SABRE substrate ($S_a$) as well as parahydrogen ($p$H$_2$, not shown) in solution bind reversibly to a reaction intermediate ($C_{1a}$) to form the main SABRE complex $C_1$; substrate dissociation and association reaction rates are denoted as $k^{d}_{1a}$ and $k^{a}_{1a}$, respectively. B) Hyperpolarized SABRE substrate ($S_a$) is coordinated with another substrate ($S_b$), usually not amenable to direct SABRE, to a second organometallic complex $C_2$; dissociation and association rates are denoted as $k^{d}_{2a}$ and $k^{a}_{2a}$, and $k^{d}_{2b}$ and $k^{a}_{2b}$ for exchange reactions involving $S_a$ and $S_b$, respectively. In the second complex, polarization can be transferred from $S_a$ to $S_b$ via spin-spin couplings.

The range of substrates amenable to SABRE generated in the direct way (in the following, simply called SABRE) is limited to molecules that can bind reversibly to Ir-based organometallic complexes used as SABRE catalysts. Thus, SABRE substrates are usually heterocycles containing an electron-donating atom — typically, nitrogen — making it possible to hyperpolarize derivatives of pyridine, purine, diazirines, Schiff bases and others.[33] The class of molecules amenable to $p$H$_2$-based hyperpolarization via SABRE can be significantly extended by using a novel method termed SABRE-relay[34-38] (the reaction schemes are shown in **Figure 1B** and **Figure 2**) recently introduced by the group of Duckett. The method utilizes either chemical exchange[35] of hyperpolarized protons between the main SABRE-active substrate $S_a$ to a second substrate $S_b$ or transfer of hyperpolarization to $S_b$ in a second complex ($C_2$)[34], which simultaneously binds $S_a$ to $S_b$ and thus, couples both their spin systems by intramolecular spin-spin couplings ($J$-couplings) in $C_2$. The SABRE-relay approach allows one to polarize substrates, which are otherwise not amenable to SABRE polarization[35] using Ir-based complexes (which bind only special classes of molecules).

In order to fully exploit the broad application potential of the SABRE-relay, we develop a consistent theoretical description of the polarization formation. The theoretical approach used in this work employs the density-matrix formalism[39], previously utilized to simulate SABRE polarization formation[40, 41]. Our theory explicitly treats the spin dynamics in the SABRE complex, the substrates and their interplay via chemical exchange. Here we limit ourselves to coherent polarization transfer at a suitable low external magnetic field, i.e., we do not consider the situation of polarization formation by incoherent polarization processes[42, 43] or high-field RF pulse sequences, since this would require consideration of the time-dependent spin Hamiltonians of $S_1$, $S_2$, $C_1$ and $C_2$. Such a generalization of the theory is possible[40] but it is beyond the scope of this work. The developed theoretical model is used for the analysis of the polarization formation process by investigating the factors affecting the resulting polarization, thus, enabling optimization of NMR signal enhancement in the SABRE-relay process.

# II. Chemical kinetic schemes

We start with a simpler approach to the problem, utilizing chemical kinetics schemes. First, we outline the derivation of the general chemical kinetics scheme[44] of the SABRE process. Subsequently, we introduce the kinetics equations of the SABRE-relay process.

## A. Chemical kinetic scheme of SABRE

Here we describe a SABRE kinetics approach, which is the first step for introducing a density matrix treatment of SABRE-relay. The consideration in this section is based on the treatment introduced by some of us previously[44] and closely follows the original formulation. In **Figure 1**, we depict the proposed SABRE kinetics, though in a simplified form. Hereafter, all concentrations are written in square brackets, i.e., [X] stands for the concentration of the compound X.

As a starting point, we write down the differential kinetic equations which arise from the exchange processes depicted in **Figure 1A**:

$$\mathrm{C_1} \leftrightarrows \mathrm{S_a} + \mathrm{C_{1a}}. \tag{1}$$

Association of substrate $S_a$ as depicted in **Figure 1A** and equation (1) describes a second-order reaction. If the concentration $[C_{1a}]$ of the intermediate $C_{1a}$ stays constant; this reaction proceeds with an effective pseudo-first-order rate constant $k_{1a}^{a} = k_{Sa}^{a}[C_{1a}]$. Additionally, here we do not explicitly describe the exchange of free hydrogen in solution with the intermediate SABRE complex $C_{1a}$, but rather assume that it is fast on NMR timescales and, thus, the spin state of the protons in $C_{1a}$ is taken the same as that of the protons in the pool of dissolved $H_2$ molecules. A treatment of the reaction intermediates is possible[36] but beyond what is required for the presented work. With these assumptions, it is possible to write down the differential equations for the concentrations of the free substrate and the main SABRE complex arising from the exchange process of equation (1):

$$\frac{d[\mathrm{C_1}]}{dt} = k_{1a}^{a}[\mathrm{S_a}] - k_{1a}^{d}[\mathrm{C_1}], \tag{2a}$$
$$\frac{d[\mathrm{S_a}]}{dt} = -k_{1a}^{a}[\mathrm{S_a}] + k_{1a}^{d}[\mathrm{C_1}]. \tag{2b}$$

The superscript indices ‘a’ and ‘d’ in equation (2) refer to the association or dissociation rate constants for the complex $C_1$, respectively. These equations are the starting point of the SABRE-relay exchange schemes described below.

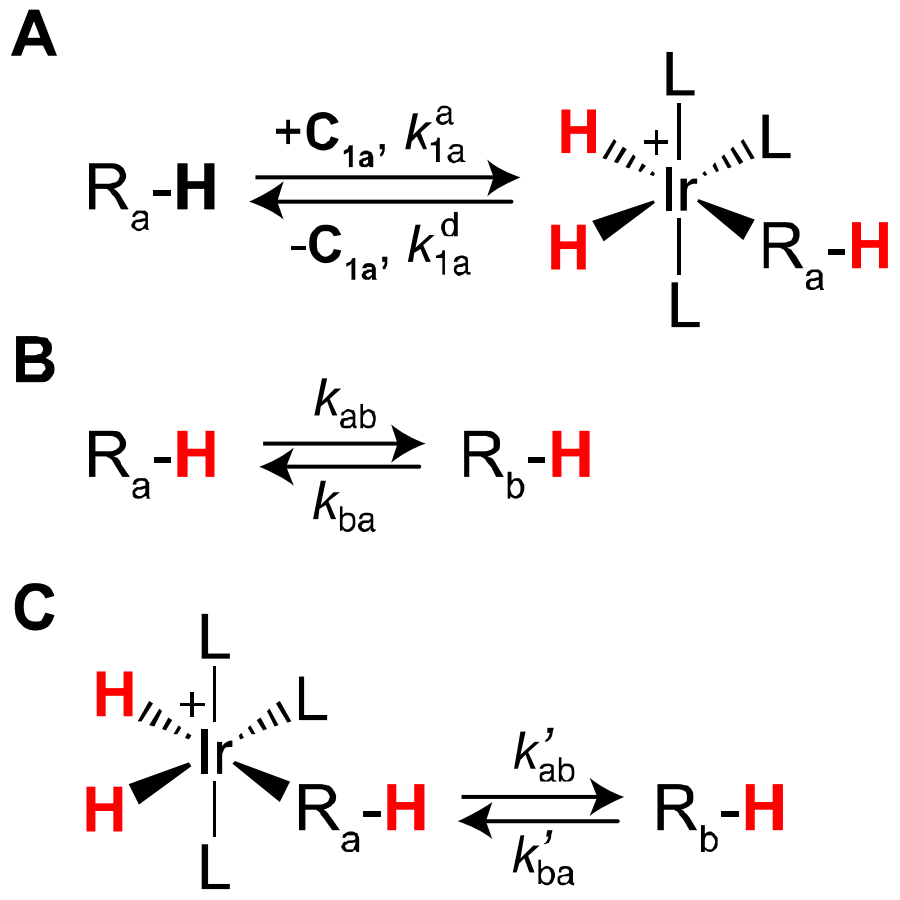

**Figure 2**: Schematic representation of the SABRE-relay process involving hydrogen exchange between two substrates (typical for amines). A) Primary SABRE-hyperpolarization process of $R_a$-H. B) Labile protons of the free substrate $R_a$-H ($S_a$) undergo chemical exchange with protons of the free substrate $R_b$-H ($S_b$), which becomes hyperpolarized in the course of the exchange process. C) Proton exchange is taking place between complex-bound substrate $R_a$-H ($S_a$) and free substrate $R_b$-H ($S_b$).

## B. Chemical kinetic scheme of SABRE-relay

In the next step, chemical exchange of hyperpolarized protons is to be considered, see **Figure 2**. Consequently, we amend the above kinetics equations by introducing the exchange reaction of the SABRE substrate $S_a$ with a second, SABRE inactive, substrate molecule $S_b$ in solution

$$S_a \leftrightarrows S_b,$$

and writing down a modified differential equation for $[S_a]$ and a new equation for $[S_b]$:

$$\frac{d[S_a]}{dt} = -k_{1a}^{a}[S_a] + k_{1a}^{d}[C_1] - k_{ab}[S_a] + k_{ba}[S_b],$$

$$\frac{d[S_b]}{dt} = k_{ab}[S_a] - k_{ba}[S_b].$$

Here $k_{ab}$ and $k_{ba}$ are reaction rates (hereafter given in $s^{-1}$) for chemical exchange of hyperpolarized nuclear spins (e.g., protons) between two substrates. These equations are similar to equation (2a), but now contain additional terms, which stand for chemical exchange between $S_a$ and $S_b$. As usual, when the system is in chemical equilibrium, there is a straightforward relation between the exchange rates and the concentrations:

$$\frac{[S_a]}{[S_b]} = \frac{k_{ab}}{k_{ba}}.$$

It should be noted that we assume a simple first-order exchange of protons between the two substrate pools. This assumption might not always hold true but can be accounted for by introducing concentration-dependent exchange rates $k_{ab}$ and $k_{ba}$.

Next, we treat a more complex situation, where a second organometallic complex $C_2$ is involved in the SABRE-relay process.[34] Such a complex binds both $S_a$ and $S_b$ (see **Figure 1B**); in the following, we extend the reaction scheme by considering coordination of $S_a$ and $S_b$ with this new complex $C_2$. To do so, we assume that the coordination of substrate ligands to $C_2$ proceeds in a manner similar to the case of the main SABRE complex. Specifically, we assume that $C_2$ dissociates into one of two reaction intermediates, $C_{2a}$ or $C_{2b}$, by dissociating $S_a$ or $S_b$, respectively. Under these conditions, we arrive at the reaction scheme given in equation (3) and depicted in **Figure 1B**:

$$\begin{aligned} \mathrm{S_a} + \mathrm{C_{2a}} &\leftrightarrows \mathrm{C_2}, \\ \mathrm{S_b} + \mathrm{C_{2b}} &\leftrightarrows \mathrm{C_2}. \end{aligned} \tag{3}$$

As has been done in Subsection II.A, we define the effective association rates $k_{2a}^{a}$ and $k_{2b}^{a}$ for $S_a$ and $S_b$ with their respective intermediate complexes $C_{2a}$ and $C_{2b}$ as

$$k_{2a}^{a} = k_{Sa}^{a}[\mathrm{C_{2a}}], \qquad k_{2b}^{a} = k_{Sb}^{a}[\mathrm{C_{2b}}].$$

As a simplifying assumption, in equation (3) we consider only complexes binding both $S_a$ and $S_b$ and do not describe complexes binding solely $S_a$ or $S_b$. Furthermore, we assume that any intermediates in this process are too short-lived to substantially alter the coherent spin dynamics in the system. The exchange process of **Figure 1B** obeys the following kinetic equations:

$$\frac{d[\mathrm{S_a}]}{dt} = -(k_{1a}^{a} + k_{2a}^{a})[\mathrm{S_a}] + k_{1a}^{d}[\mathrm{C_1}] + k_{2a}^{d}[\mathrm{C_2}],$$

$$\frac{d[\mathrm{S_b}]}{dt} = -k_{2b}^{a}[\mathrm{S_b}] + k_{2b}^{d}[\mathrm{C_2}],$$

$$\frac{d[\mathrm{C_2}]}{dt} = k_{2a}^{a}[\mathrm{S_a}] + k_{2b}^{a}[\mathrm{S_b}] - \left(k_{2a}^{d} + k_{2b}^{d}\right)[\mathrm{C_2}].$$

In Section III, we introduce the equations for the density matrices of the substrates and complexes involved. Such equations describe both the spin dynamics and chemical kinetics, as well as their interplay.

# III. Density matrix calculations

Here we generalize the density matrix approach, which has been previously used to describe the SABRE process, to the more complex situation of SABRE-relay. Specifically, we introduce the density-matrix equations for the species $S_b$, to which hyperpolarization is relayed, and, if needed, for the SABRE-relay complex $C_2$. The density matrix operations, which are used to describe association and dissociation (direct product and partial trace), are the same as those used before[41] and are explained in more detail in Subsection III.C.

## A. SABRE-relay via chemical exchange

In the case of SABRE-relay via chemical exchange, it becomes necessary to introduce the density matrix for the species $S_b$ and consider chemical exchange between $S_a$ and $S_b$. Here we use a simplified spin system, namely, we assume that the molecules of species $S_a$ and $S_b$ contain only a single proton, which is exchanging with the forward and backwards rates $k_{ab}$ and $k_{ba}$, respectively (**Figure 2A**). The set of equations for the three density matrices of interest (density matrices of the two substrates, $\hat{\sigma}_{S_a}$ and $\hat{\sigma}_{S_b}$ and complex $\hat{\sigma}_{C_1}$) is as follows:

$$\begin{aligned} \frac{d}{dt}\hat{\sigma}_{S_a} &= \hat{\hat{L}}_{S_a}\hat{\sigma}_{S_a} + k_{1a}^{d}\mathrm{Tr}_{C_{1a}}\{\hat{\sigma}_{C_1}\} - k_{1a}^{a}\hat{\sigma}_{S_a} + k_{ba}\hat{\sigma}_{S_b} - k_{ab}\hat{\sigma}_{S_a}, \\ \frac{d}{dt}\hat{\sigma}_{C_1} &= \hat{\hat{L}}_{C_1}\hat{\sigma}_{C_1} - k_{1a}^{d}\hat{\sigma}_{C_1} + k_{1a}^{a}\{\hat{\rho}_{H_2} \otimes \hat{\sigma}_{S_a}\}, \end{aligned} \tag{4}$$

$$\frac{d}{dt}\hat{\sigma}_{\mathrm{S_b}} = \hat{\hat{L}}_{\mathrm{S_b}}\hat{\sigma}_{\mathrm{S_b}} - k_{\mathrm{ba}}\hat{\sigma}_{\mathrm{S_b}} + k_{\mathrm{ab}}\hat{\sigma}_{\mathrm{S_a}}.$$

The system of differential equations (4) describes the exchange of protons between the species $S_a$ and $S_b$ only when they both are in the free form. Here $\hat{\hat{L}}_{\mathrm{X}} = -i\hat{\hat{H}}_{\mathrm{X}} + \hat{\hat{R}}_{\mathrm{X}}$ is the Liouville operator of the species X, where $\hat{\hat{H}}$ is the commutator-defined superoperator of the spin Hamiltonian, $\hat{\hat{H}}_{\mathrm{X}}\hat{\sigma}_{\mathrm{X}} = [\hat{H}_{\mathrm{X}}, \hat{\sigma}_{\mathrm{X}}]$, and $\hat{\hat{R}}_{\mathrm{X}}$ is the relaxation superoperator. The precise form of the Hamiltonians $\hat{H}_{\mathrm{X}}$ is specified in the Subsection III.C; we always assume that the conditions for efficient coherent polarization transfer with respect to the magnetic field are achieved, meaning the system is at the proper level anti-crossing (LAC) conditions.[20, 21]

Simple arguments on the way of writing density matrix equations (4) are the following. As a starting point, we use the corresponding kinetic equations for the concentrations and replace $[\mathrm{X}] \rightarrow \hat{\sigma}_{\mathrm{X}}$ in all equations, i.e., replace the concentration by the corresponding density matrix, which is normalized as $\mathrm{Tr}\{\hat{\sigma}_{\mathrm{X}}\} = [\mathrm{X}]$. There are only two additional issues which have to be taken into account: (i) terms describing the nuclear spin evolution $\hat{\hat{L}}_{\mathrm{X}}\hat{\sigma}_{\mathrm{X}}$ should be added and (ii) the dimensionality of the matrices on the left-hand side and on the right-hand side should be matched. To do so, when necessary, we reduce dimensionality by the partial trace operation (when substrate is dissociated from the complex) and increase dimensionality by taking the direct product, $\otimes$, of density matrices (when substrate and the complex are associated). These simple considerations are in agreement with the rigorous derivation of kinetic equations, which can be found elsewhere.[45, 46]

In the particular case under consideration, equation (4), the terms describing substrate exchange with the complex are introduced in the same was as previously.[41] To reduce the dimensionality of the term describing dissociation in the equation for $\hat{\sigma}_{\mathrm{S_a}}$ we take the partial trace over the spin states of hydrides in the complex $C_1$. To increase the dimensionality of the density matrix in the term, which describes association in the equation for $\hat{\sigma}_{\mathrm{C_1}}$, we take the direct product of $\hat{\sigma}_{\mathrm{C_{1a}}} = \hat{\rho}_{\mathrm{H_2}}$ and $\hat{\sigma}_{\mathrm{S_a}}$, where $\hat{\rho}_{\mathrm{H_2}} = |S\rangle\langle S|$ is the normalized spin density matrix of $H_2$. It is given by the projection operator onto the singlet state $|S\rangle$ (in contrast to $\hat{\sigma}_{\mathrm{X}}$ matrices, the trace of the density matrix $\hat{\rho}_{\mathrm{H_2}}$ is equal to unity). The terms describing proton exchange are introduced by using the rate constants $k_{\mathrm{ab}}$ and $k_{\mathrm{ba}}$.

So far, we only considered the situation where the protons of $S_a$ and $S_b$ exchange, assuming that both substrates are in their free form in solution, i.e., when $S_a$ is not bound to the SABRE complex $C_1$. However, the situation may arise, where the protons of the catalyst-bound species $S_a$ exchange with those of $S_b$, see **Figure 2C**. Indeed, chemical exchange involving complex-bound substrates in SABRE has been reported before in the case of coordinated water.[47, 48]

We can account for this process, by adding appropriate terms to equation (4):

$$\begin{aligned}
\frac{d}{dt}\hat{\sigma}_{\mathrm{S_a}} &= \hat{\hat{L}}_{\mathrm{S_a}}\hat{\sigma}_{\mathrm{S_a}} + k_{\mathrm{1a}}^{\mathrm{d}}\mathrm{Tr}_{\mathrm{C_{1a}}}\left(\hat{\sigma}_{\mathrm{C_1}}\right) - k_{\mathrm{1a}}^{\mathrm{a}}\hat{\sigma}_{\mathrm{S_a}} + k_{\mathrm{ba}}\hat{\sigma}_{\mathrm{S_b}} - k_{\mathrm{ab}}\hat{\sigma}_{\mathrm{S_a}} \\
\frac{d}{dt}\hat{\sigma}_{\mathrm{C_1}} &= \hat{\hat{L}}_{\mathrm{C_1}}\hat{\sigma}_{\mathrm{C_1}} - k_{\mathrm{1a}}^{\mathrm{d}}\hat{\sigma}_{\mathrm{C_1}} + k_{\mathrm{1a}}^{\mathrm{a}}\left(\hat{\rho}_{\mathrm{H_2}} \otimes \hat{\sigma}_{\mathrm{S_a}}\right) - k'_{\mathrm{ab}}\hat{\sigma}_{\mathrm{C_1}} + k'_{\mathrm{ba}}\left(\mathrm{Tr}_{\mathrm{S_a}}\left(\hat{\sigma}_{\mathrm{C_1}}\right) \otimes \hat{\sigma}_{\mathrm{S_b}}\right) \\
\frac{d}{dt}\hat{\sigma}_{\mathrm{S_b}} &= \hat{\hat{L}}_{\mathrm{S_b}}\hat{\sigma}_{\mathrm{S_b}} - \left(k_{\mathrm{ba}} + k'_{\mathrm{ba}}\right)\hat{\sigma}_{\mathrm{S_b}} + k_{\mathrm{ab}}\hat{\sigma}_{\mathrm{S_a}} + k'_{\mathrm{ab}}\mathrm{Tr}_{\mathrm{C_{1a}}}\left(\hat{\sigma}_{\mathrm{C_1}}\right)
\end{aligned} \qquad (5)$$

For simplicity, we assume here that this process also occurs with the first-order rate constants $k'_{\mathrm{ab}}$ for forward and $k'_{\mathrm{ba}}$ for reverse exchange, respectively. Here we again assume that any other reaction intermediates are too short-lived to alter the spin dynamics.

## B. SABRE-relay in a second complex

Next, we formulate the set of equations suitable to describe the relayed transfer of polarization in a second SABRE-relay complex (see **Figure 3**). Such a complex is treated in a similar way as the main SABRE complex, in the sense that it temporarily binds both $S_a$ and $S_b$. While both substrates reside at the catalyst, the spins of both $S_a$ and $S_b$ are connected by scalar spin-spin couplings, *J*-couplings, allowing for coherent transfer of spin order. The system of equations used in this work is written as follows:

$$
\begin{aligned}
\frac{d}{dt}\hat{\sigma}_{S_a} &= \hat{\hat{L}}_{S_a}\hat{\sigma}_{S_a} + k_{1a}^{d}\,\mathrm{Tr}_{C_{1a}}\{\hat{\sigma}_{C_1}\} - (k_{1a}^{a} + k_{2a}^{a})\hat{\sigma}_{S_a} + k_{2a}^{d}\mathrm{Tr}_{S_b}\{\hat{\sigma}_{C_2}\} \\
\frac{d}{dt}\hat{\sigma}_{C_1} &= \hat{\hat{L}}_{C_1}\hat{\sigma}_{C_1} - k_{1a}^{d}\,\hat{\sigma}_{C_1} + k_{1a}^{a}\{\hat{\rho}_{H_2}\otimes\hat{\sigma}_{S_a}\} \\
\frac{d}{dt}\hat{\sigma}_{S_b} &= \hat{\hat{L}}_{S_b}\,\hat{\sigma}_{b} + k_{2b}^{d}\,\mathrm{Tr}_{S_a}\{\hat{\sigma}_{C_2}\} - k_{2b}^{a}\hat{\sigma}_{S_b} \\
\frac{d}{dt}\hat{\sigma}_{C_2} &= \hat{\hat{L}}_{C_2}\hat{\sigma}_{C_2} - \left(k_{2a}^{d} + k_{2b}^{d}\right)\hat{\sigma}_{C_2} + k_{2a}^{a}\left\{\hat{\sigma}_{S_a}\otimes \mathrm{Tr}_{S_a}\{\hat{\sigma}_{C_2}\}\right\} + k_{2b}^{a}\{\mathrm{Tr}_{b}\{\hat{\sigma}_{C_2}\}\otimes\hat{\sigma}_{b}\}
\end{aligned}
\tag{6}
$$

Here we treat the spin dynamics for all chemical species involved, i.e., the substrates $S_a$ and $S_b$ and complexes $C_1$ and $C_2$. As previously, we introduce the spin evolution of each species by using the corresponding Liouville operator; chemical exchange is introduced in the same way as above (when necessary, the dimensionality of the density matrices is reduced by taking partial trace or increased by taking direct product as described in subsection III.C).

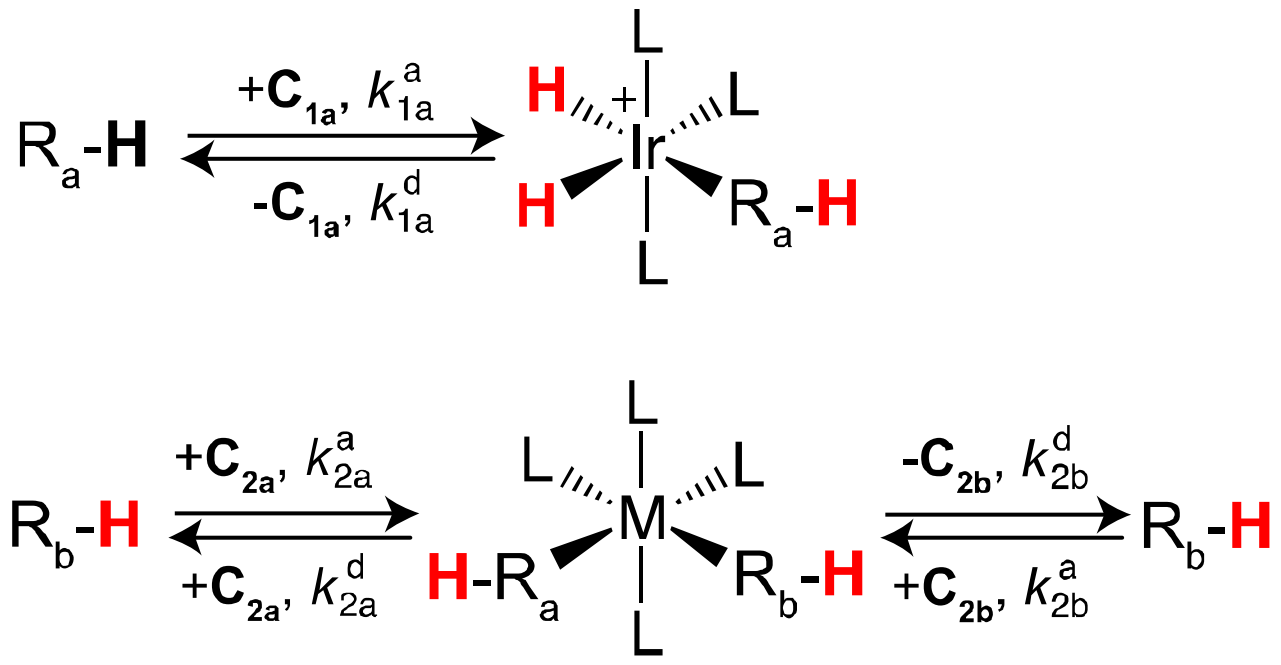

**Figure 3**: Schematic representation of SABRE-relay process involving chemical exchange between two organometallic complexes. Note that polarization transfer to protons is shown while other nuclei in the second complex can be polarized as well under optimal magnetic field conditions.

## C. Implementation of the model

Using sets of equations (4) and (6), one can treat the reaction and spin dynamics for both schemes of the SABRE-relay experiment. However, we still need to comment on implementation of the model, explaining how different superoperators should be introduced in the numerical scheme and how the set of equations should be solved. We also comment on the NMR observables discussed in the rest of the paper.

In contrast to the density matrix formulations previously used to describe SABRE,[41] the above systems of equations are non-linear, which precludes using a simple linear propagation operator to solve them. Here we chose to either integrate the system numerically (time traces can be found in the **Appendix A**) using the Runge-Kutta method or to obtain the steady-state solution by solving the equation systems using the Levenberg-Marquardt method. When integration was used, the system was evolved to a time point of at least five times the $T_1$-relaxation time of the slowest relaxing spin species in order to guarantee that a steady-state of the hyperpolarization built-up has been reached. All calculations were performed on a standard office laptop with 2 cores. Each integration of the system took between 1 and 30 seconds. When solving the equation system to obtain the steady-state solution using the Levenberg-Marquardt method, the calculation time was approximately an order of magnitude longer. We assume that initially all spins in the system are non-polarized, except for the $p\mathrm{H_2}$-nascent protons.

The spin dynamics are governed by the Liouville operators of the individual species introduced in equation (6)

$$\hat{\hat{L}}_{\mathrm{X}} = -i\hat{\hat{H}}_{\mathrm{X}} + \hat{\hat{R}}_{\mathrm{X}}$$

Here the spin Hamiltonians of each species, comprising $N$ spins, are written as follows.

$$\hat{\mathcal{H}} = \sum_{i=1}^{N} \omega_i \hat{I}_{i,z} + \sum_{i<j} J_{ij}(\hat{\mathbf{I}}_1 \cdot \hat{\mathbf{I}}_2)$$

Here $\omega_i = -\gamma B_z(1+\delta_i)$ where $\gamma_i$ and $\delta_i$ are the gyromagnetic ratio and chemical shift of the $i$-th spin. In case of numerical integration and dealing with only one spin species (namely, protons), we neglect the large Larmor frequency in order to speed up calculations and just retain the chemical shift part of the Zeeman term taking $\omega_i = -\gamma_H B_z \delta_i$. When the steady state solution was obtained or X-nuclei were considered, the full Zeeman term was retained. To model relaxation, we employ a previously described treatment of random fluctuating fields.[49] The parameters used (unless otherwise stated in the Section IV) can be found in **Table 1**. Considering spin relaxation, we always use a homogeneous term $\hat{\hat{R}}_{\mathrm{X}}\hat{\sigma}_{\mathrm{X}}$ rather than $\hat{\hat{R}}_{\mathrm{X}}(\hat{\sigma}_{\mathrm{X}} - \hat{\sigma}_{\mathrm{X}}^{\mathrm{eq}})$, thus neglecting the small equilibrium spin polarization described by the density matrix $\hat{\sigma}_{\mathrm{X}}^{\mathrm{eq}}$.

We assume all species to be in stationary conditions; thus the association rates in equation (6) can be expressed in terms of the dissociation rate and the concentrations of the species in the system. More precisely, if we consider exchange between two species A and B with concentrations [A] and [B] as well as forward and reverse rates $k_{\mathrm{A}\to\mathrm{B}}$ and $k_{\mathrm{B}\to\mathrm{A}}$, we always check that the relation $\frac{k_{\mathrm{A}\to\mathrm{B}}}{k_{\mathrm{B}\to\mathrm{A}}} = \frac{[\mathrm{B}]}{[\mathrm{A}]}$ is fulfilled. This is needed to make sure that the trace of the individual density matrices and hence the concentrations in the system do not change during the calculations. In **Appendix B**, we briefly reiterate the density matrix treatment used to calculate the effects of exchange here. The parameters governing the spin-dynamics, i.e. J-couplings and chemical shifts, as well as the exchange rates of different species are also listed in **Appendix B**.

For interpretation of the simulation results reported below, we calculate the polarization of the spins in different species in the following way:

$$P(\mathrm{X}) = 2 \cdot \frac{\mathrm{Tr}\{\hat{\sigma}_{\mathrm{X}} \cdot \hat{I}_z\}}{\mathrm{Tr}\{\hat{\sigma}_{\mathrm{X}}\}}$$

Additionally, in some cases we are interested to calculate NMR signal which is proportional to magnetization, the product of spin polarization and concentration of the corresponding species X:

$$M_{\mathrm{eff}}(\mathrm{X}) = P(\mathrm{X}) \cdot [\mathrm{X}]$$

# IV. Results and Discussion

## A. SABRE-relay via chemical exchange

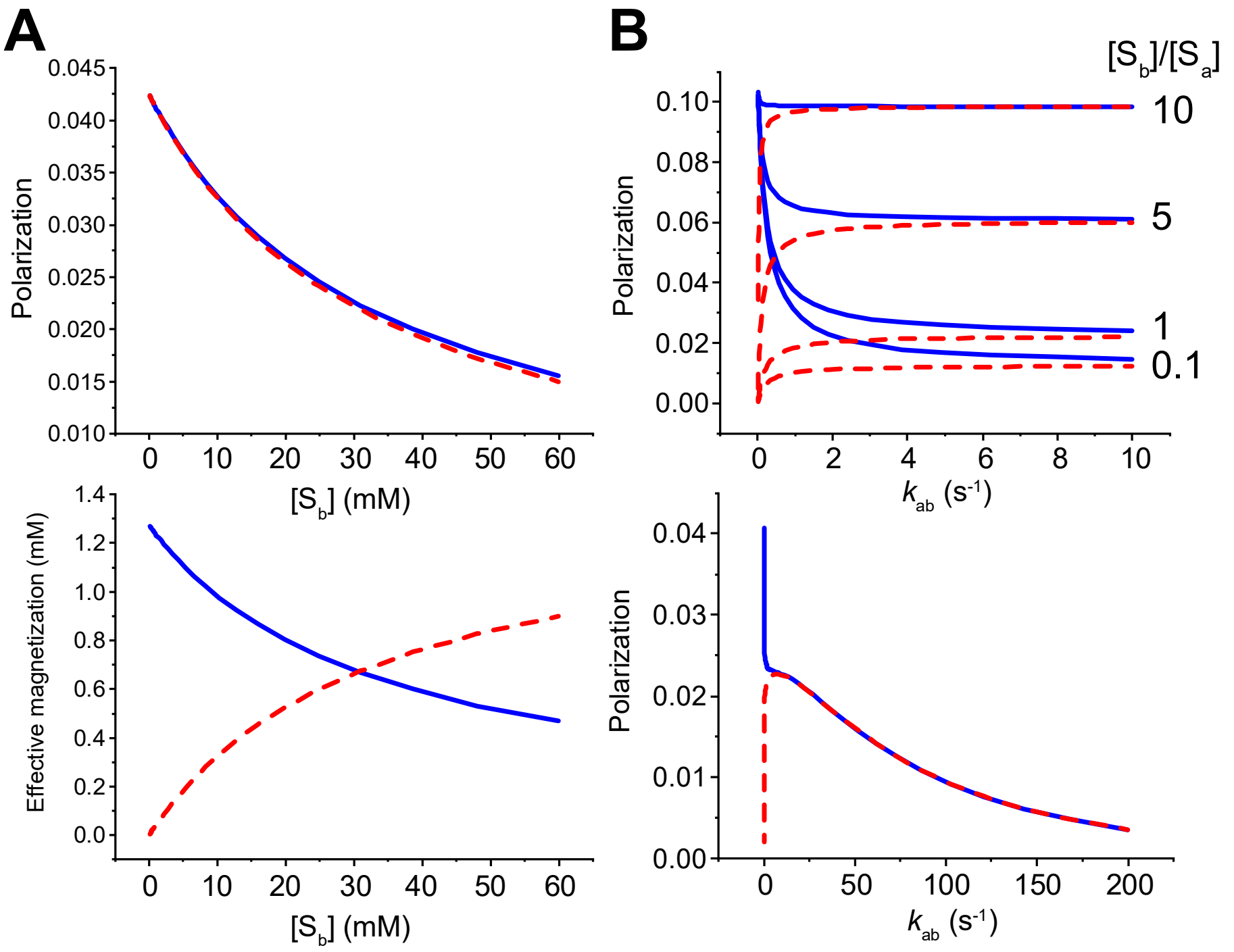

Figure 4: (A) Polarization (top) and effective magnetization (bottom) of the main SABRE substrate ($S_a$, blue line) and the SABRE-relay substrate ($S_b$, red dashed line) as a function of the concentration $[S_b]$ while $[S_a]$ is kept constant. (B) Top – polarization of the main SABRE substrate ($S_a$, blue line) and SABRE-relay substrate ($S_b$, red dashed line), as a function of the exchange rate $k_{ab}$ for different ratios $[S_b]/[S_a]$ (top). Here $[S_a]$ = 10 mM is kept constant. Bottom – polarization of the main SABRE substrate ($S_a$, blue line) and SABRE-relay substrate ($S_b$, red dashed line), depending on the exchange rate $k_{ab}$ when free $S_a$ and free $S_b$ are exchanging. Here we account for both processes depicted in Figure 2B and 2C with kinetic rate constants $k'_{ab} = k_{ab}$. Note that the process depicted in Figure 2C interferes with the coherent polarization process of $S_a$ in $C_1$ at high values of $k'_{ab}$.

First, we examine the polarization levels for the second species $S_b$ considering the SABRE-relay model of chemical exchange $S_a \rightleftarrows S_b$, see **Figure 2A,B**, as described by the set of equations (4). When both the main SABRE substrate $S_a$ as well as the second substrate $S_b$ undergo fast chemical exchange, the polarization of both will decrease as the concentration of either one is increased (see **Figure 4A**). Such a behaviour has been previously predicted for the concentration dependence of the main SABRE substrate[41, 44] and can be explained by the fact that only a limited amount of the substrate can be polarized by the main SABRE complex per unit time. On the other hand, relaxation of the free substrate pool drives its polarization back to the small equilibrium value. Thus, at a certain concentration, the overall number of hyperpolarized spins will be the same, regardless of the size of the free substrate pools. Consequently, the average polarization decreases as the pool is increased. This effect can be better understood considering what happens to the effective magnetization (see previous section) in the system, which is shown in **Figure 2A,B**. When the concentration of species $S_b$ is low, so is its signal. As the concentration of $S_b$ increases (we assume that $S_b$ undergoes rapid exchange with species $S_a$), more and more of the hyperpolarized spins will be found in the $S_b$ pool, rather than in the $S_a$ pool. Consequently, the signal of $S_b$ will increase and eventually plateau, while the signal of $S_a$ decreases. However, a distinction to previously obtained results for the concentration dependence of SABRE polarization should be made: in the SABRE approach, the substrate molecules reduce the hyperpolarization efficiency by competing with $pH_2$ (the source of hyperpolarization) in the exchange with the main SABRE complex. Such a behaviour is not occurring here for $S_b$, as the $S_b$ molecules never bind to any complex.

Next, we want to explore the polarization dependence on the exchange rate of protons between $S_a$ and $S_b$, which is plotted in **Figure 4B**. One can see that at low exchange rates the second substrate is not polarized at all, while with an increasing rate of exchange, the polarization is distributed between both species $S_a$ and $S_b$. The total polarization in this case depends on the relative concentrations (i.e., a larger pool of substrate again leads to a lower polarization value).

So far, exchange only between the two substrates in their free form was considered. If, however, the SABRE-relay substrate $S_b$ is also exchanging (here we assume for simplicity the rates are the same as for the free form) with the bound form of $S_a$ in the main SABRE complex, as depicted in **Figure 2C**, the spin evolution changes significantly. In such a situation, the exchange between the protons of the two substrates interferes destructively with the coherent polarization transfer mechanism in the main SABRE complex. This is because a certain residence time at the catalyst is needed for effective coherent transfer of polarization from $pH_2$ to the substrate protons of $S_a$.[19, 41, 44] Consequently, when the proton exchange rates are too high, the efficiency of SABRE-relay drops significantly (see **Figure 4B**). We speculate, that this behaviour might be the reason for the lower efficiency of the relayed SABRE polarization of amines reported in the presence of water in the sample.[37]

## B. SABRE-relay in a second complex

Let us now turn to the situation where polarization is relayed in a second organometallic complex.

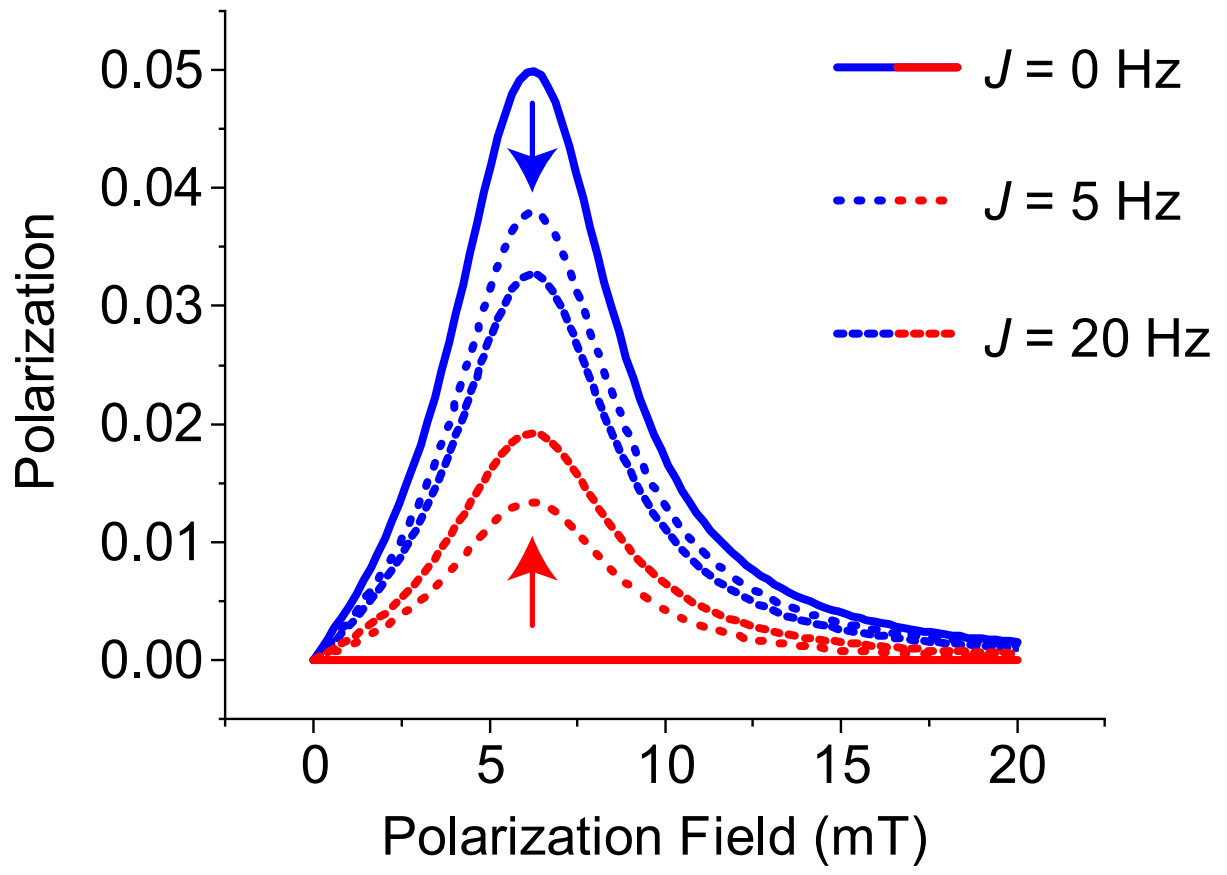

**Figure 5**: Polarization Field dependence of the main SABRE substrate ($S_a$, blue) and SABRE-relay substrate ($S_b$, red), depending on the *J*-couplings in the SABRE-relay. Here a chemical shift difference of 10 ppm between $S_a$ and $S_b$ was assumed.

**Polarization transfer mechanism**. As mentioned before, in the scheme with two complexes the transfer is not mediated by chemical exchange of protons, but by coherent transfer of polarization via *J*-couplings in the second complex (although chemical exchange of protons between ligands bound to the catalyst, as reported before[47, 48], could also be treated with the above equations in a straightforward way). The efficiency of spin polarization transfer between the two nuclei belonging to $S_a$ and $S_b$ will thus depend on the difference $\Delta\nu = \gamma_H B_z \Delta_\delta / 2\pi$ in their Zeeman interactions with the field frequency difference and *J*-couplings (here $\Delta_\delta$ is the chemical shift difference). **Figure 5** shows the field dependence for a system of two indirectly coupled protons in the SABRE-relay complex. It should be noted that because of this small polarization field, already relatively small couplings can be efficient to transfer polarization between protons, even if their chemical shift difference is large (here 10 ppm were assumed which at 5 mT corresponds to a frequency difference of approximately 2.5 Hz). However, for transfer to most X-nuclei, this process will become efficient only at the appropriate ultralow field, where SABRE polarization of protons is then again inefficient[4]. Hence, we see that the magnetic field strength, favourable for spin order transfer $pH_2 \rightarrow S_a$, is also suitable for the relayed polarization transfer $S_a \rightarrow S_b$ in the second complex when protons are considered.

**Concentration dependences of polarization**. When examining the predicted polarization dependence of the two substrates $S_a$ and $S_b$ on the concentration of the main SABRE complex $C_1$ as well as of the SABRE-relay complex $C_2$, a curious dependence is found. While an increase of $C_1$ leads to an increasing polarization of both substrates, eventually reaching a maximum, increase of the concentration of the SABRE-relay complex results gives rise to a different behaviour (see **Figure 6A**). At small $[C_2]$, the increase of the concentration of the SABRE-relay complex leads to an increase of the polarization of both substrates. Upon further increase, however, $S_a$ and $S_b$ equilibrate and decrease together. This behaviour is explained by the relatively fast relaxation ($R = 1\ s^{-1}$) of the substrates bound to the organometallic SABRE complexes, as assumed here. Thus, when the concentration of these complexes is increased, the effective $T_1$-relaxation time of the substrates, and consequently their polarization, is reduced. However, for the main SABRE complex $C_1$, this reduction (caused by enhanced $T_1$-relaxation) is compensated by an increased production of hyperpolarized species in the system.

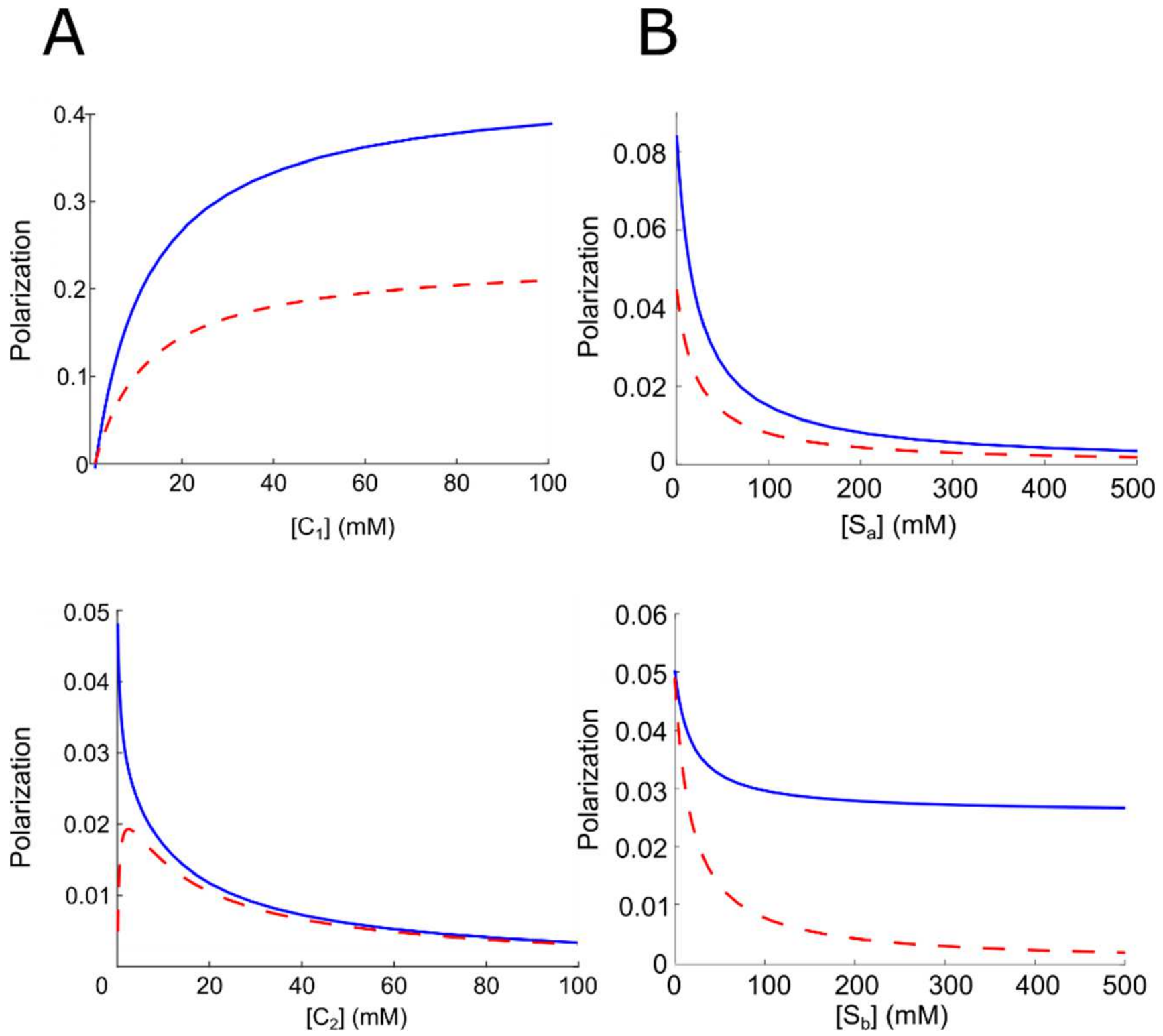

**Figure 6**: (A) Polarization as a function of the concentration the main SABRE complex $C_1$ (top) and SABRE-relay complex $C_2$ (bottom); here the results are shown for substrate $S_a$ (blue line) and substrate $S_b$ (red dashed line). While increase of the $[C_1]$ concentration leads to an increase and eventual levelling-off of the polarization of both substrates, increase of $[C_2]$ gives rise to a monotonous decrease of the polarization of $S_1$ and a dependence with a maximum for the polarization of $S_b$. (B) Polarization dependence of the main SABRE substrate ($S_a$, blue line) and the relay substrate ($S_b$, red dashed line) on the concentrations of the free substrate $S_a$ (top) and $S_b$ (bottom).

The dependence of hyperpolarization of the free substrates $S_a$ and $S_b$ on their concentrations, shown in **Figure 6B**, can be rationalized in the following way. When the $[S_a]$ concentration is increased, its polarization drops, similar to the results in the previous subsection and as predicted by previously formulated SABRE models.[41, 44] Consequently, because $S_a$ acts as the source of polarization distributed into the second complex, the polarization of $S_b$ also decreases. When the concentration of $S_b$ is varied, the behaviour is somewhat different. Upon increase of $[S_b]$, its hyperpolarization decreases, reflecting the fact that the amount of hyperpolarized molecules per unit of time, at best, is independent of the free substrate pool and its increase will not increase the hyperpolarized magnetization generated in the system. The polarization of $S_a$ will reach a constant value which is independent of the amount of SABRE-relay complexes in the system, because at high concentrations of $S_b$, the amount of molecules hyperpolarized by relayed transfer from $S_b$ will stay constant.

**Dependence of polarization on exchange rate**. We examined the dependence of the hyperpolarization generated by SABRE-relay not only on concentrations, but also on the kinetic parameters, namely, on the exchange rates in complex $C_2$. For the coherent polarization transfer mechanism considered here, our investigation predicts an optimal dissociation rate in the second complex (see **Figure 7A**). This behaviour is similar to that predicted[19, 41] for SABRE, namely, at high exchange rates (and consequently, short lifetimes of the complex) the coherent transfer of polarization is suppressed, whereas at low exchange rates the generated polarization is limited by relaxation.

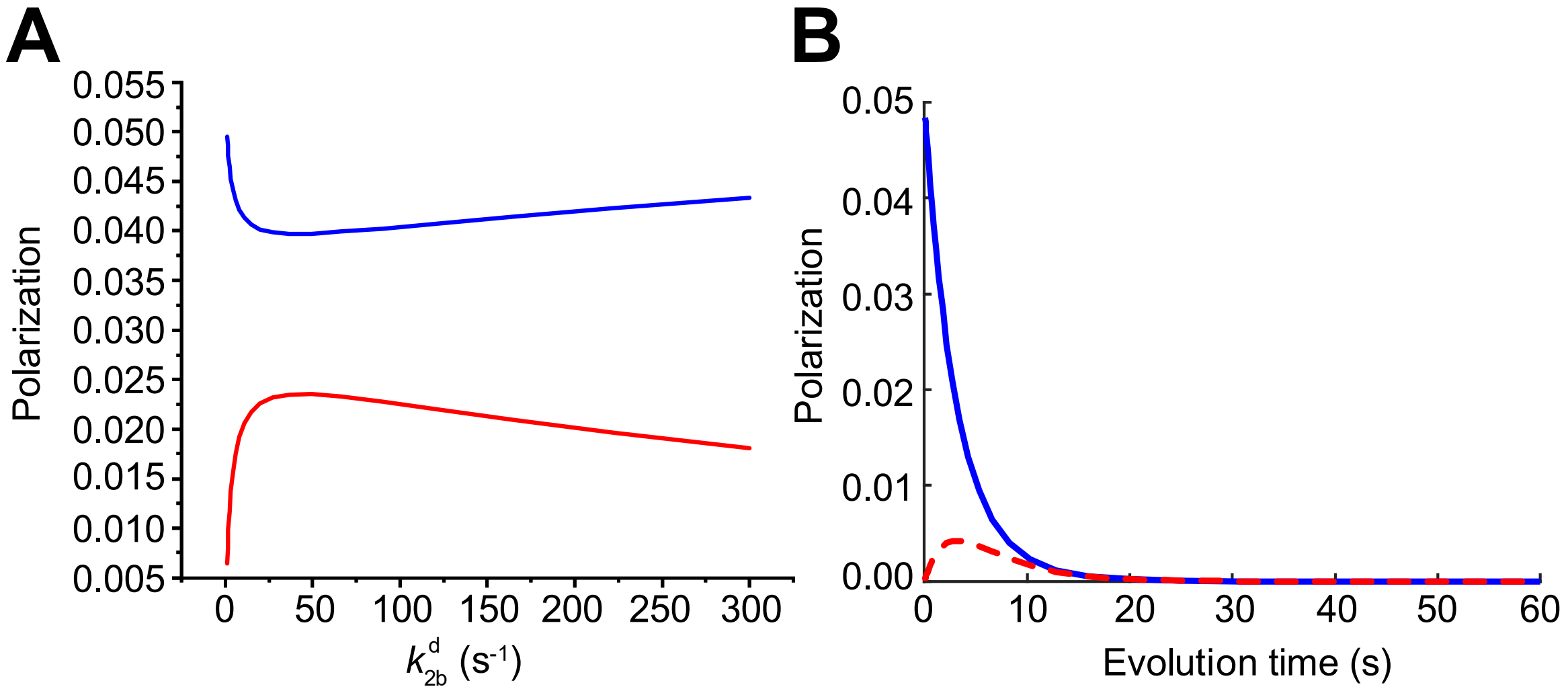

**Figure 7**: (A) Polarization of the main SABRE substrate $S_a$ (blue line) and the relay substrate $S_b$ (red dashed line) as functions of the dissociation rate ($k^{d}_{2b}$) of the SABRE-relay complex. (B) Temporal evolution of polarization (ensemble-averaged quantity) of the proton of the main SABRE substrate $S_a$ (blue line) and a $^{31}P$ nucleus in the second complex $C_2$ (red dashed line). In the calculation we assumed initial polarization of protons at the optimal SABRE field, 6 mT with a subsequent evolution at an ultralow field, here, at 1 µT.

**Relayed transfer of polarization to heteronuclei**. As discussed above, relayed polarization transfer can occur not only among protons, but also among protons and heteronuclei. As an example of the versatility of our model to simulate SABRE-relay experiments, we calculate the time-dependence of polarization transferred to a $^{31}P$ nucleus in the second organometallic complex $C_2$ as described by Roy and co-workers[34]. The simulated polarization scheme is as follows: in the first stage, the primary SABRE substrate $S_a$ is polarized at a field of 5 mT. Consequently, the field is lowered to a value, which correspond to strong coupling between $^{31}P$ and $^{1}H$ nuclei. Ошибка! Источник ссылки не найден.**7B** shows the evolution of polarization at this second field, which is 1 µT. The simulations demonstrate that transfer of proton hyperpolarization of the SABRE substrate to the $^{31}P$ nuclei in the second complex reported[34] can be reasonably well reproduced by the developed model.

# V. Conclusion and Outlook

To summarize, in this work we present the first theoretical model to describe the emerging hyperpolarization method SABRE-relay. A detailed analysis of the SABRE-relay efficiency on both the spin degrees of freedom (*J*-couplings and NMR frequencies) as well as on the concentrations and exchange rates of the chemical constituents of this system was conducted in order to guide future development of this field. As SABRE-relay has been shown to be applicable to a stunning number of systems, we do not aim to provide a full description of all possible formulations and applications of the presented theoretical approach but rather aim at laying the groundwork for future developments.

By using the proposed method, one can analyse the dependence of polarization on kinetic and spin parameters of the system under consideration, which is the crucial step for understanding the efficiency of the SABRE-relay approach and for optimizing its performance. We anticipate that the present treatment can support development of SABRE-relay and its extension to a broad range of substrates, which cannot be polarized by the traditional SABRE method.

# Acknowledgements

K.L.I. acknowledges support from the Russian Science Foundation (project 19-43-04116) and G.B. acknowledges financial support by the Deutsche Forschungsgemeinschaft under contract Bu-911-29-1.

# Appendices

## Appendix A: Time dependence of polarization

In this Appendix, we provide exemplary time traces for the numerical integration of the two different SABRE relay approaches described above.

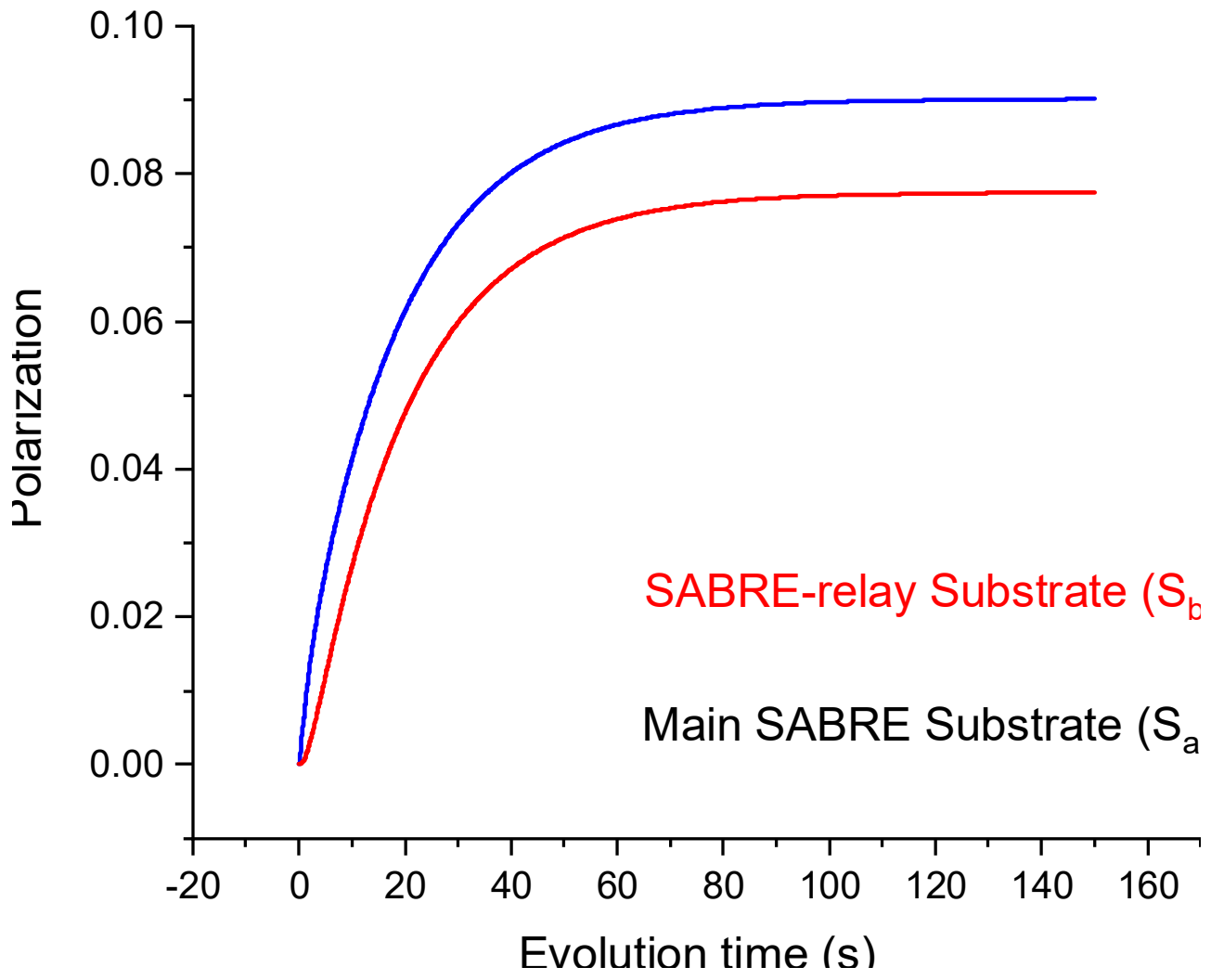

**Figure A1**: Temporal evolution of polarization of the proton of the main SABRE substrate ($S_a$, blue) and the SABRE-relay substrate ($S_b$, red). Here $T_1$ of both substrates was 30s and the J-coupling connecting them was assumed to be 10 Hz.

Representative time traces for both SABRE-relay mechanisms are shown in **Figures A1** and **A2**. In both cases, the behaviour of polarization is qualitatively the same: primarily $S_a$ is polarized in the first SABRE complex. At later times, polarization is transferred to $S_b$ either by chemical exchange or by polarization transfer in the second complex. Polarization of $S_b$ is thus build-up at later times and to a lower level. In the analysis presented in the main part of the paper, we present only the steady-state solutions for polarization, achieved at $t \to \infty$.

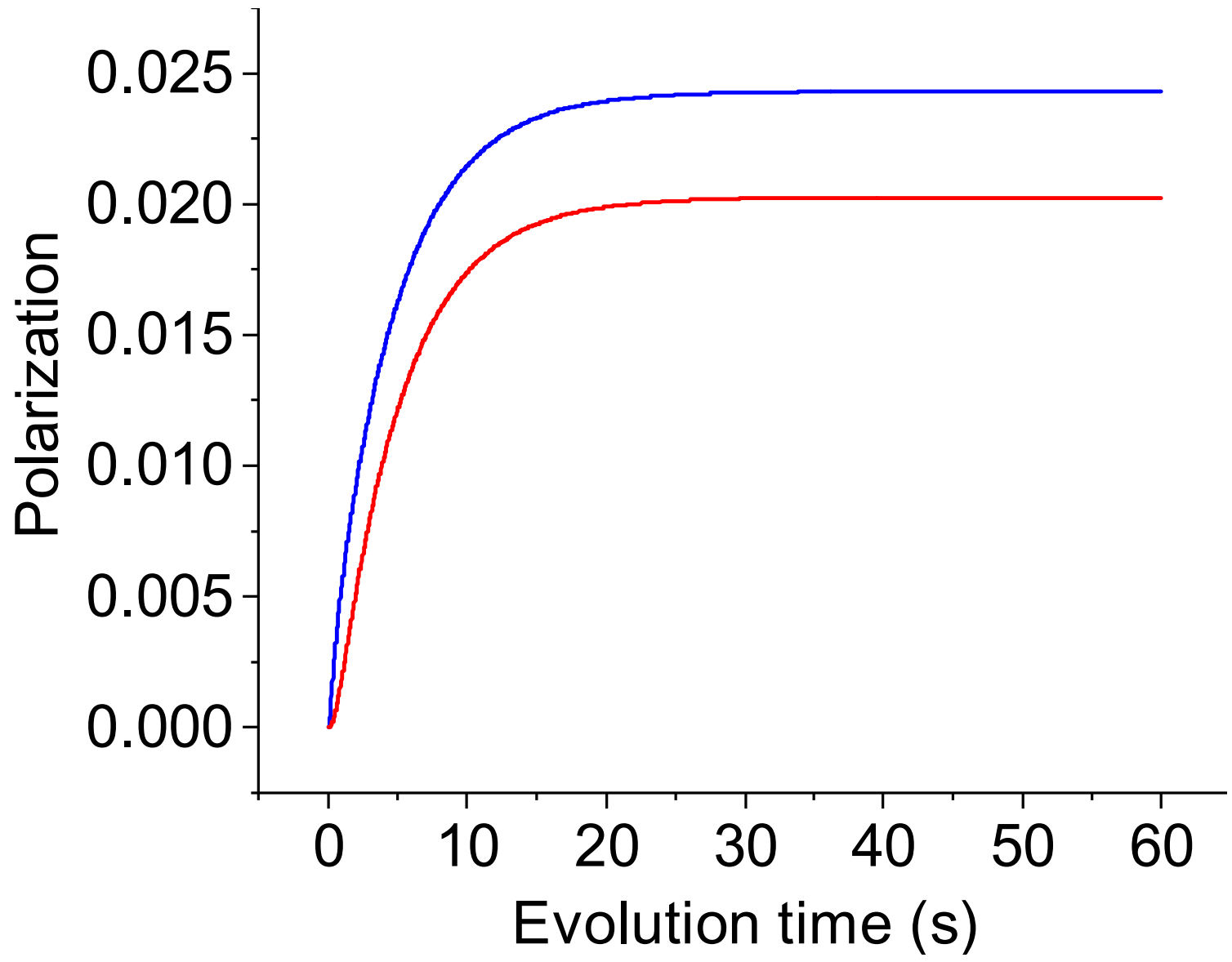

**Figure A2**: Temporal evolution of polarization of the proton of the main SABRE substrate ($S_{1a}$, blue) and the SABRE-relay substrate ($S_b$, red). Here, the chemical exchange rate $k_{ab}$ was 1s.

## Appendix B: details of the simulation

In the following, we briefly reiterate the density matrix treatment used to calculate the effects of exchange here. This approach has been described in detail previously and we refer the interested reader to the appropriate literature.[41] Let us again assume that we are dealing with two species A and B with concentrations [A] and [B] whose spin system is described by two normalized density matrices $\hat{\rho}_{\mathrm{A}}$ and $\hat{\rho}_{\mathrm{B}}$. First, we choose to normalize our density matrices by their concentrations:

$$\hat{\sigma}_{\mathrm{A}} = \hat{\rho}_{\mathrm{A}}[\mathrm{A}], \qquad \hat{\sigma}_{B} = \hat{\rho}_{\mathrm{B}}[\mathrm{B}]$$

Under such normalization the trace of each density matrix is proportional to the corresponding concentration. If we assume that these two species coordinate to form a complex

$$A + B \leftrightarrow C$$

we need to introduce appropriate terms in the equations for the density matrices. When the spins of A and B are coupled in the complex C, they should be described by a common density matrix $\hat{\sigma}_C$. If this reaction proceeds with a rate constant $k^{\mathrm{a}}$, then association contributes to the differential equation governing the dynamics of $\hat{\sigma}_C$ in the following way:

$$\left\{\frac{d}{dt}\hat{\sigma}_{\mathrm{C}}\right\}_{ass} = k^{\mathrm{a}}\{\hat{\sigma}_{\mathrm{A}} \otimes \hat{\sigma}_{\mathrm{B}}\}$$

If the complex C dissociates again into its components A and B, we treat this by assuming (justified by the random nature of exchange in a large ensemble)[45] that all coherences between A and B, which may have existed in the complex, are lost. Thus, such a dissociation process, governed by a dissociation rate $k_{\mathrm{d}}$, contributes to the differential equations of $\hat{\sigma}_{\mathrm{A}}$ and $\hat{\sigma}_{\mathrm{B}}$ in the following way:

$$\left\{\frac{d}{dt}\hat{\sigma}_{\mathrm{A}}\right\}_{diss} = \mathrm{Tr}_{\mathrm{B}}\{\hat{\sigma}_{\mathrm{C}}\}, \qquad \left\{\frac{d}{dt}\hat{\sigma}_{\mathrm{B}}\right\}_{diss} = \mathrm{Tr}_{\mathrm{A}}\{\hat{\sigma}_{\mathrm{C}}\}$$

Where $\mathrm{Tr}_{\mathrm{X}}$ is the partial trace operation over the states of the spin system X.

The parameters of the spin system used in this work, unless stated otherwise, can be found in **Table 1**. Accordingly, the concentrations and exchange rates are summarized in **Table 2**.

**Table 1**. Parameters of the spin systems used in calculations, *J*-couplings in the two SABRE complexes, chemical shifts and relaxation rates ($R = 1/T_1 = 1/T_2$). All nuclei are protons.

| | *J* (Hz) | | | | $R_1$ ($s^{-1}$) | $\delta$ (ppm) |
|---|---|---|---|---|---|---|
| SABRE complex | | $H_1$ | $H_2$ | $S_a$ | | |
| $H_1$ | | | –7.7 | | 1 | –22 |
| $H_2$ | | –7.7 | | | 1 | –22 |
| $S_a$ | | 0 | 1 | | 0.3 | 8.3 |
| SABRE-relay complex | | $S_1$ | $S_2$ | | | |
| $S_a$ | | | 1 | | 1 | 8.3 |
| $S_b$ | | 1 | | | 1 | 8.3 |
| Free SABRE substrate | | | | | | |
| $S_a$ | | | | | 0.2 | 8 |
| SABRE Relay Substrate | | | | | | |
| $S_a$ | | | | | 0.2 | 8 |

**Table 2**. Chemical parameters used in calculations.

| Species | Concentrations (mM) | Exchange rates ($s^{-1}$) |
|---|---|---|
| SABRE complex | 1 | $k_{ab} = 10$ |
| SABRE-relay complex | 1 | $k_d = 20$ |
| $S_a$ | 30 | |
| $S_b$ | 30 | |